% ****** Start of file apssamp.tex ******
\documentclass[%
 reprint,
 amsmath,amssymb,
 aps,
]{revtex4-1}

\usepackage{graphicx}% Include figure files
\usepackage{dcolumn}% Align table columns on decimal point
\usepackage{bm}% bold math 
\usepackage{color}
\usepackage{hyperref}% add hypertext capabilities
\usepackage{ulem}
\begin{document}

\preprint{APS/123-QED}

\title{Neutrino Mass Ordering - Circumventing the Challenges using Synergy between T2HK and JUNO}

\author{Sandhya Choubey}
\email[]{choubey@kth.se}
\affiliation{Department of Physics, School of Engineering Sciences, KTH Royal Institute of Technology, AlbaNova University Center, Roslagstullsbacken 21, SE--106~91 Stockholm, Sweden}
\affiliation{The Oskar Klein Centre for Cosmoparticle Physics, AlbaNova University Center, Roslagstullsbacken 21, SE--106 91 Stockholm, Sweden}

\author{Monojit Ghosh}
\email[]{monojit$\_$rfp@uohyd.ac.in}
\affiliation{School of Physics, University of Hyderabad, Hyderabad - 500046, India}
\affiliation{Center of Excellence for Advanced Materials and Sensing Devices, Ruder Bo\v{s}kovi\'c Institute, 10000 Zagreb, Croatia}

\author{Deepak Raikwal}
\email[]{deepakraikwal@hri.res.in}
\affiliation{Harish-Chandra Research Institute,  A CI of Homi Bhabha National Institute, Chhatnag Road, Jhunsi, Prayagraj - 211019, India}
\affiliation{Homi Bhabha National Institute, Anushakti Nagar, Mumbai 400094, India}

\date{\today}% It is always \today, today,

\begin{abstract}
One of the major open problems of neutrino physics is MO (mass ordering). We discuss the prospects of measuring MO with two under-construction experiments T2HK and JUNO. JUNO alone is expected to measure MO with greater than $3\sigma$ significance as long as certain experimental challenges are met. In particular, JUNO needs better than 3\% energy resolution for MO measurement. On the other hand, T2HK has rather poor prospects at measuring the MO, especially for certain ranges of the CP violating parameter $\delta_{\rm CP}$, posing a major drawback for T2HK. In this article we show that the synergy between JUNO and T2HK will bring two-fold advantage. Firstly, the synergy between the two experiments helps us determine the MO at a very high significance. With the baseline set-up of the two experiments, we have a greater than $9\sigma$ determination of the MO for all values of $\delta_{\rm CP}$. Secondly, the synergy also allows us to relax the constraints on the two experiments. We show that JUNO, could perform extremely well even for energy resolution of 5\%, while for T2HK the MO problem with ``bad" values of $\delta_{\rm CP}$ goes away. The MO sensitivity for the combined analysis is expected to be greater than $6\sigma$ for all values of $\delta_{\rm CP}$ and with just 5\% energy resolution for JUNO. 

\end{abstract}

\maketitle

%\tableofcontents

\section{\label{sec:level1}Introduction}

Despite the big strides made in the field of neutrino physics, we still lack knowledge on some of the important neutrino parameters. The neutrino mass ordering (MO) is a prime example and measuring it will have far-reaching consequences both theoretically as well as experimentally. By MO we essentially mean the structure of the neutrino mass spectrum. For three generations of neutrinos, one can define two mass-squared differences, which we call $\Delta m_{21}^2$ and $\Delta m_{31}^2$ ($\Delta m_{ij}^2=m_i^2 - m_j^2$). While it is experimentally known that $\Delta m_{21}^2 > 0$, the sign of $\Delta m_{31}^2$ is still not statistically confirmed. The case with $\Delta m_{31}^2>0$ is referred to as the normal mass ordering (NO), while $\Delta m_{31}^2<0$ is called inverted mass ordering (IO) \cite{Esteban:2020cvm}. 

Experiments are being built that would be able to shed light on MO. Such experiments are expensive as well as well as challenging. The JUNO experiment \cite{JUNO:2015zny} is being built in China specifically to determine the MO \cite{IceCube-Gen2:2019fet,Cao:2020ans,KM3NeT:2021rkn,Forero:2021lax}. JUNO would observe electron antineutrinos from powerful Chinese reactors and must reach an energy resolution of at least 3\% or better in order to achieve a modest statistical significance. This is a major technological challenge and so-far unprecedented. On the other hand, the T2HK experiment \cite{Abe:2016ero} under construction will be observing muon type and electron type neutrinos (and antineutrinos) from accelerator facilities in Japan \cite{Ghosh:2017ged,Raut:2017dbh,Cho:2019ctv,Agarwalla:2017nld,Fukasawa:2016yue,Chatterjee:2017xkb,Blennow:2014sja,King:2020ydu,Chakraborty:2017ccm,C:2014mmz,Ghosh:2019sfi,Coloma:2012wq,Ballett:2016daj,Evslin:2015pya,Panda:2022vdw}. This experiment has measurement of the CP violating phase $\delta_{\rm CP}$ as its major goal. However, the unresolved MO is a major challenge for this experiment that leads to a deterioration of its sensitivity. In fact T2HK's MO sensitivity is rather low, particularly for $\delta_{\rm CP}=0^\circ$. In this article we will study the synergy between these two experiments and show for the first time that combining the data from just these two experiments will (i) determine MO to very high significance (ii) reduce the challenges they face. In particular, we will show that JUNO would be able to perform remarkably well for energy resolutions of even 5\%, for which even the current technology suffices. This will bring down both the technological demand as well as costs on JUNO. 

In this article we will focus on the role of $|\Delta m_{31}^2|$ in the MO determination. Data corresponding to NO can be mimicked (or fitted) by a theory corresponding to IO by changing the value of $|\Delta m_{31}^2|$ in the fit. We will show, both analytically as well as numerically, how different neutrino oscillation channels change $|\Delta m_{31}^2|$ differently in the fit. This change is also dependent on the neutrino energy, therefore, different neutrino energy bins would require different amount of shift in $|\Delta m_{31}^2|$. Since the experiments depend on different oscillation channels and since we also have spectral information in these experiments, this leads to a spectacular synergy between different experiments, as well as between different data bins of the same experiment. 

In another study~\cite{Raikwal:2022nqk} we will include the future atmospheric neutrino experiment ICAL at the INO facility~\cite{ICAL:2015stm} and will be focusing on how its sensitivity is affected due to synergy with JUNO and T2HK. However, the main focus of the present work is two-fold. Firstly, we study quantitatively how much the synergy between JUNO and T2HK improves the MO sensitivity of the combined set-up. Secondly, and even more importantly, we study how the synergy helps in reducing the energy resolution requirement on JUNO. We will quantitatively show that even if JUNO is unable to reach its design energy resolution of 3\%, we would still be able to measure the MO with high significance. Therefore, the two papers are complementary to each other.

\section{\label{sec:level2}The Method}

The expected MO sensitivity for these future facilities is estimated by the following method. The prospective data is calculated for these experiments assuming NO. This simulated data is then statistically fitted with a theory of IO by defining a $\chi^2$. We simulate both T2HK and JUNO using the GLoBES software \cite{Huber:2004ka,Huber:2007ji}. For details of the experimental parameters used in this work we refer to \cite{Abe:2016ero} for T2HK and \cite{JUNO:2015zny} for JUNO. For T2HK we have two kinds of data sets. The data on muon (and antimuon) events depend on the $\nu_\mu \to \nu_\mu$ survival probability $P_{\mu\mu}$ which is referred to as the ``disappearance channel". On the other hand, the data on electron (and positron) events depend on the $\nu_\mu \to \nu_e $ conversion probability $P_{\mu e}$ and is referred to as the ``appearance channel". JUNO has only data set for positrons which depends on the $\bar\nu_e \to \bar\nu_e$ survival probability $P_{\bar e\bar e}$. 

The ``data" for T2HK and JUNO are simulated for NO at the values of the oscillation parameters given in column 2 of Table \ref{Tab:Param}. We fit the ``data" with IO and allow $\theta_{23}$, $\delta_{\rm CP}$ and $|\Delta m_{31}^2|$ to vary in the range given in column 3 of Table \ref{Tab:Param}. The parameters $\theta_{12}$, $\theta_{13}$ and $\Delta m^2_{21}$ are kept fixed in the fit at their values given in column 2. 
\begin{table}[h]
	\centering
	\setlength{\extrarowheight}{0.1cm}
	\begin{tabular}{|c|c|c|}
		\hline
		 Parameter & Best-fit value & $3 \sigma$ range \\
		\hline
		$\theta_{12}$ & $33.44^\circ$ & - \\
		$\theta_{13}$ & $8.57^\circ$ & - \\
		$\theta_{23}$ & $45^\circ$ & $40^\circ$ to $52^\circ$ \\
		$\delta_{\rm CP}$ & $-90^\circ/0^\circ$ & $-180^\circ$ to $180^\circ$  \\
		$\Delta m^2_{21}$ & $7.42\times10^{-5}$ eV$^2$ & -  \\
		$\Delta m^2_{31}$ &  $2.531\times10^{-3}$ eV$^2$ & (2.435 to 2.598) $\times 10^{-3}$ eV$^2$ \\
		\hline
	\end{tabular}
	\caption{The best-fit values and $3 \sigma$ ranges of the oscillation parameters used in our calculation.}
	\label{Tab:Param}
\end{table}

\section{\label{sec:dm31}The pivotal role of $|\Delta m^{2}_{31}|$}

For both T2HK and JUNO matter effects are negligible and the survival probability of $\nu_\alpha$, where $\alpha=e$ or $\mu$, can be approximately given as
\begin{eqnarray} \nonumber
P_{\alpha\alpha}&=& 1-4|U_{\alpha 1}|^{2}|U_{\alpha 2}|^{2}\sin^{2}\Delta_{21}-4|U_{\alpha 1}|^{2}|U_{\alpha 3}|^{2}\sin^{2}\Delta_{31} \\
&-& 4|U_{\alpha 2}|^{2}|U_{\alpha 3}|^{2}\sin^{2}\Delta_{32},
\label{eq:psurv}
\end{eqnarray}
where $\Delta_{ij}=\Delta m_{ij}^{2}L/4E$, $L$ being the distance that the neutrino travels and $E$ its energy. This gives $P_{\alpha\alpha}^{NO}$ for $\Delta_{31}^{\rm NO}$ and $P_{\alpha\alpha}^{\rm IO}$ for $\Delta_{31}^{\rm IO}$ and one can obtain an analytic expression for $\Delta P_{\alpha\alpha}=P_{\alpha\alpha}^{\rm NO}-P_{\alpha\alpha}^{\rm IO}$. Note that since $\Delta m^{2}_{32} = \Delta m^{2}_{31} - \Delta m^{2}_{21}$, and since we keep $\Delta m^{2}_{21}>0$ fixed, we have 
\begin{eqnarray} \nonumber
&&\Delta P_{\alpha\alpha}= -4|U_{\alpha 3}|^{2}\bigg[ |U_{\alpha 1}|^{2}\bigg(\sin^{2}\Delta_{31}^{\rm NO} - \sin^{2}\Delta_{31}^{\rm IO} \bigg)\\
&-& |U_{\alpha 2}|^{2}\bigg(\sin^{2}(\Delta_{31}^{\rm NO} - \Delta_{21}) - 
\sin^{2}(\Delta_{31}^{\rm IO} + \Delta_{21})\bigg)\bigg],
\label{eq:delpaa}
\end{eqnarray}
The Eq.~(\ref{eq:delpaa}) shows that $\Delta P_{\alpha\alpha}$ cannot be zero for $|\Delta_{31}^{\rm NO}|=|\Delta_{31}^{\rm IO}|$ since the second term  becomes non-zero for this case. This term will reduce in magnitude if $|\Delta_{31}^{\rm IO}|$ is reduced, however, then the first term becomes non-zero. Note that the first term is weighted by $|U_{\alpha 1}|^2$, while the second term is weighted by $|U_{\alpha 2}|^2$. Therefore, the minimum of $\Delta P_{\alpha\alpha}$ comes at a value of $|\Delta_{31}^{\rm IO}|$ that is lower than $|\Delta_{31}^{\rm NO}|$ roughly by $\Delta m^{2}_{21}$ weighted by a factor that depends on $|U_{\alpha 1}|^2$ and $|U_{\alpha 2}|^2$. Therefore, the following points are evident. Since the mass ordering sensitivity is given in terms $\Delta P_{\alpha\alpha}$, if the data corresponds to NO, then the fit will change the value of $|\Delta_{31}^{\rm IO}|$ such that $\Delta P_{\alpha\alpha}$ is minimized. Also, if $\Delta_{31}^{\rm IO}=-\Delta_{31}^{\rm NO}+x$, then $x$ will be different for $P_{e e}$ and $P_{\mu\mu}$ since it depends on $|U_{\alpha 1}|^2$ and $|U_{\alpha 2}|^2$. Therefore, (1) the best fit value of $\Delta m_{31}^2$ is different for T2HK and JUNO leading to synergy between them. Also, since the discussion above involves $\Delta_{31} = \Delta m_{31}^2 L/4E$, the minima of $\Delta P_{\alpha\alpha}$ will appear at  different values of $\Delta m_{31}^2$(IO) for different values of $E$, (2) even within the same experiment, the different $E$ bins will have a different best-fit $\Delta m_{31}^2$(IO), leading to synergy even within the same experiment. Finally, (3) T2HK also has the appearance channel $P_{\mu e}$ which gives best-fit $\Delta m_{31}^2$(IO) at a different value leading to further synergy. We will show how the synergy between $P_{\mu\mu}$ and $P_{\mu e}$ also leads to increase in MO sensitivity of T2HK. We show how the three above mentioned synergies lead to an increase of MO sensitivity.

\subsubsection{T2HK}

For the disappearance channel, following the discussion above we note that close to the $\Delta m_{31}^2$ oscillation maximum, the minima for $\Delta P_{\mu\mu}$ appears when \cite{Nunokawa:2005nx,Raut:2012dm} 
\begin{equation}
x=\frac{2|U_{\mu 2}|^{2}}{|U_{\mu 1}|^{2}+|U_{\mu 2}|^{2}}\Delta m_{21}^{2}
\label{eq:valx}
\end{equation}

\begin{figure}[h]
\hspace{-0.38 in}
    \centering
    \includegraphics[width=0.5\textwidth]{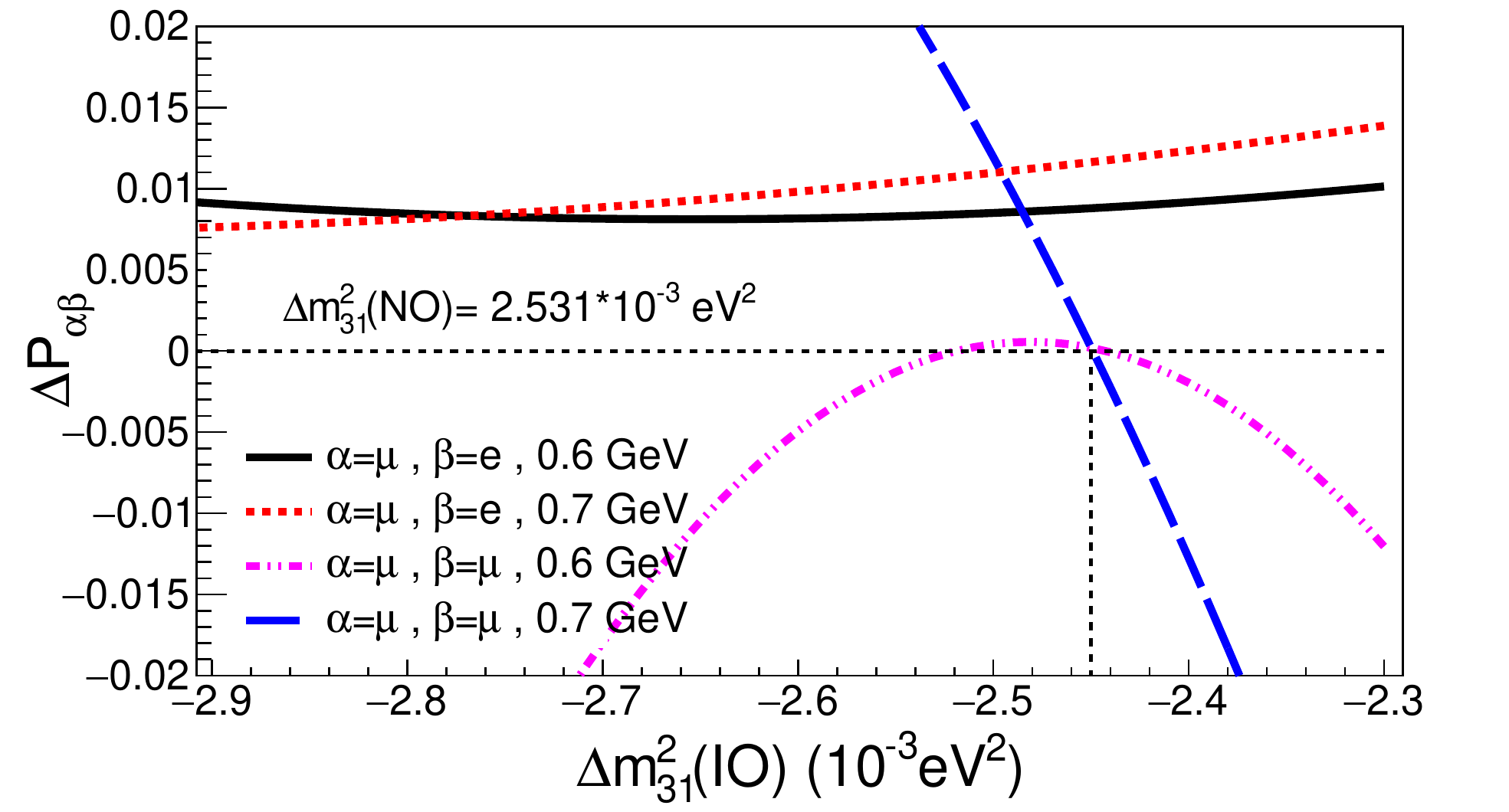}
    \caption{$\Delta P_{\mu\mu}$ and $\Delta P_{\mu e}$ as a function of  $\Delta m_{31}^{2}$(IO) for $L=295$ km and two values of energy.  }
    \label{fig:probt2hk}
\end{figure}

For our choice of oscillation parameters, we obtain $x=0.0813\times10^{-3}$ eV$^2$, giving 
$\Delta m_{31}^{2}(IO) =(-2.531+0.0813)\times10^{-3} ~\rm{eV}^{2}  = 
-2.4497\times10^{-3} ~{\rm {eV}}^{2}$. We show $\Delta P_{\mu\mu}$ in Fig.~\ref{fig:probt2hk} for $L=295$ km and two neutrino energies $E=0.6$ and $0.7$ GeV. This figure has been obtained from a numerical calculation of the full three-generation oscillation probability including earth matter effects. 

Note that for the disappearance channel, NO can be matched almost exactly by IO for $\Delta m_{31}^{2}(\rm IO) \approx -2.4497\times10^{-3} ~{\rm {eV}}^{2}$, which agrees well with Eq.~(\ref{eq:valx}). Note that we get $\Delta P_{\mu\mu}\simeq 0$ at almost the same value of $\Delta m_{31}^{2}(\rm IO)$ for both energies since they are both close to the oscillation maxima. The figure also shows the appearance channel. Since the appearance channel has more matter effects, $\Delta P_{\mu e}$ is never zero and the best-fit $\Delta m_{31}^{2}(\rm IO)$ comes at a different value.

\begin{figure}
\hspace{-0.38 in}
    \centering
    \includegraphics[width=0.5\textwidth]{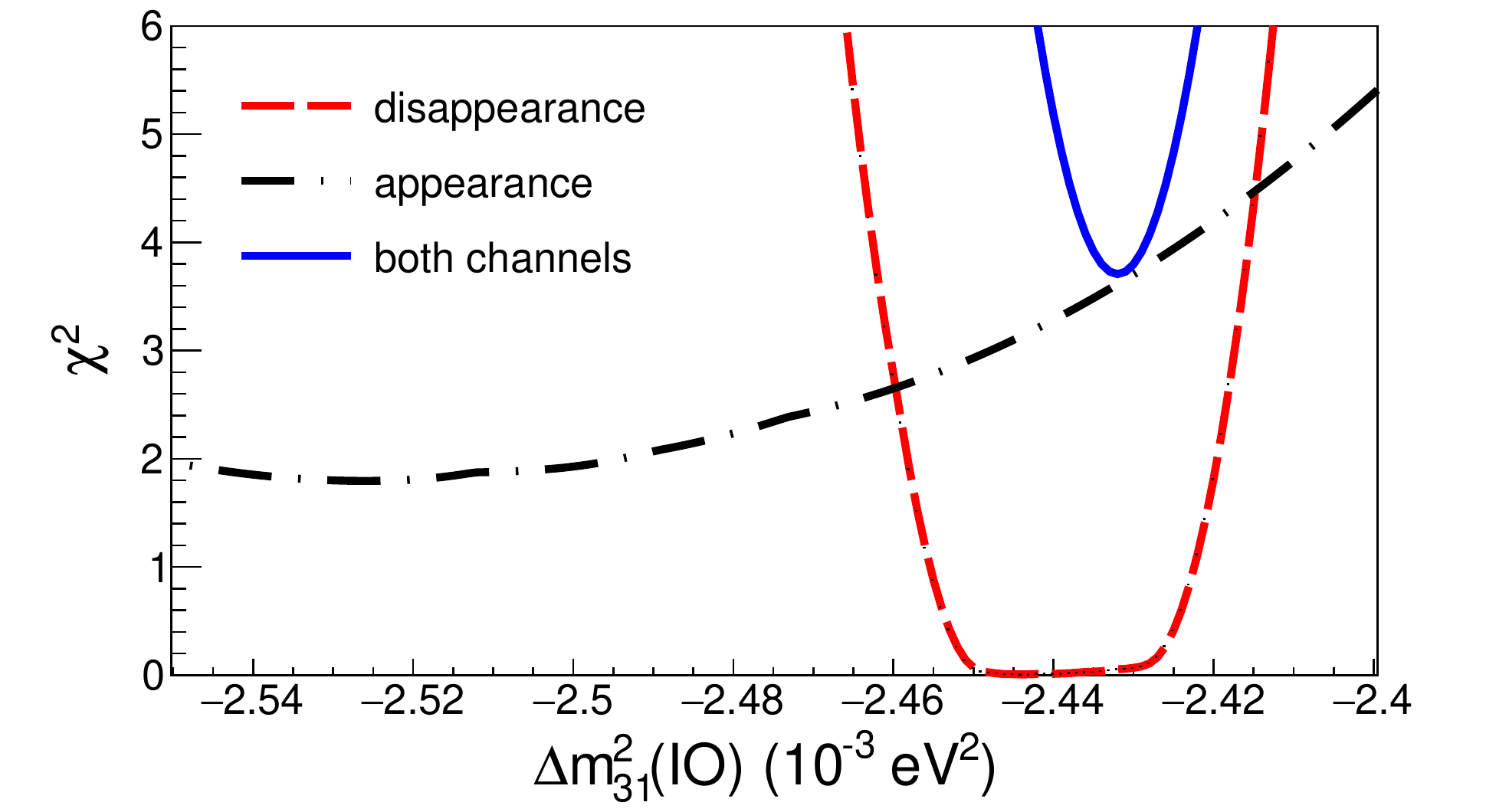}
    \caption{Mass ordering sensitivity $\chi^{2}$ as a function of  $\Delta m_{31}^2$(IO) for T2HK.}
    \label{prob_comb}
\end{figure}

In Fig.~\ref{prob_comb} we show the effect of combining the disappearance and the appearance channels in T2HK in a $\chi^2$ plot. One can see that the minima in $\Delta m_{31}^2$(IO) for the disappearance channel comes very close to that predicted in Eq.~(\ref{eq:valx}) and Fig.~\ref{fig:probt2hk} and different as compared to the assumed true value. For the appearance channel, the  $\Delta m_{31}^2$(IO) dependence of the $\chi^2$ is shallow in comparison, nevertheless, it too has a distinct minimum, which is not only different from the assumed true value, but also different with the minimum for the disappearance channel. As a result, when we combine them (blue line) we get a significantly higher $\chi^2$. 

\subsubsection{JUNO}

The Eqs.~(\ref{eq:psurv}) and (\ref{eq:delpaa}) are also valid for JUNO and 
studies on JUNO along these lines have been performed before \cite{Stanco:2017bpq}. Here we propose a somewhat novel approach. Note that Eq.~(\ref{eq:delpaa}) is given in terms of $\Delta_{31}^{\rm IO}$ which depends on both $\Delta m_{31}^2$(IO) and $E$. Since T2HK is a narrow band beam peaked at its oscillation maximum it was enough to work at $E$ corresponding to oscillation maximum. However, JUNO is a wide band beam and $E$ plays an important role here. This essentially means that for each $E$ we will have a different value of $\Delta m_{31}^2$(IO) that will give $\Delta P_{\bar e \bar e}=0$. So we compute analytically and find solutions for 
\begin{equation}
    \frac{\partial (\Delta P_{\bar e \bar e})}{\partial \Delta_{31}^{\rm IO}} = 0,
\end{equation}
giving us the following relation
\begin{eqnarray}
&& \cos^2\theta_{12}\sin 2\Delta_{31}^{\rm IO}+\sin^2\theta_{12}\sin2(\Delta_{31}^{\rm IO}-\Delta_{21}) \nonumber \\ \nonumber
&-&\cos^2\theta_{12}\sin 2\Delta _{31}^{\rm NO}.\frac{\Delta_{31}^{\rm NO}}{\Delta_{31}^{\rm IO}}-\sin^2\theta_{12}\sin2\Delta_{32}^{\rm NO}\frac{\Delta_{32}^{\rm NO}}{\Delta_{31}^{\rm IO}}=0    .
\label{eq:minrelation}
\end{eqnarray}

\begin{figure}
\hspace{-0.38 in}
    \centering
    \includegraphics[width=0.5\textwidth]{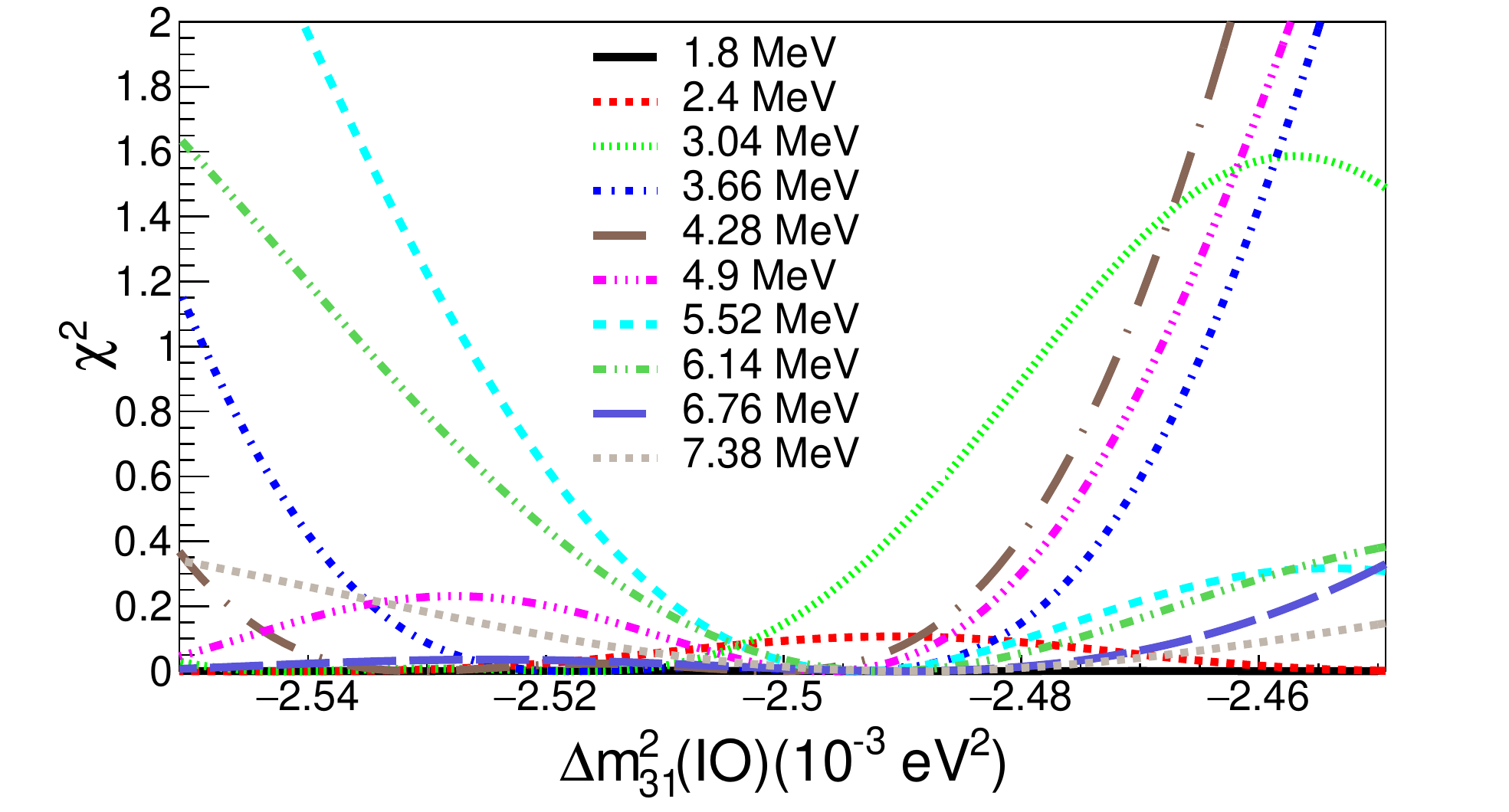}
    \caption{Mass ordering $\chi^2$ in JUNO for 10 energy bins out of the total 200 energy bins. Systematic uncertainties are neglected.}
    \label{fig:junomin}
\end{figure}

This is a relation between $E$ and $\Delta m_{31}^2$(IO) which gives us a minimum in $\Delta P_{\bar e \bar e} $. The reactor antineutrino event spectrum that we consider for JUNO varies between $E=1.8$ MeV to 8 MeV and is divided into 200 equispaced bins, with bin width 0.031 MeV. For the sake of illustration, we compute the mass ordering $\chi^2$ for each bin in JUNO, neglecting all systematic uncertainties and show these for 10 example bins in Fig.~\ref{fig:junomin}. 
We note two important points from this figure. Firstly, the different $E$ bins give $\chi^2$ minimum at different values of $\Delta m_{31}^2$(IO). Secondly, and even more importantly, note that $\chi^2_{min} \simeq 0 $ in each of the individual bins, albiet at a different value of $\Delta m_{31}^2$(IO). Hence, when we do the combined analysis of all the 200 bins, we get $\chi^2 = 10$. This is due to the synergy between the different $E$ bins of JUNO as described above. 
The best-fit $\Delta m_{31}^2$(IO) obtained from the full 200 bin analysis of JUNO is different from the assumed true value of $\Delta m_{31}^2$(NO) used in the simulated data.

\section{Combined sensitivity of JUNO and T2HK}

\begin{figure*}
\parbox{8cm}{
\includegraphics[width=9cm]{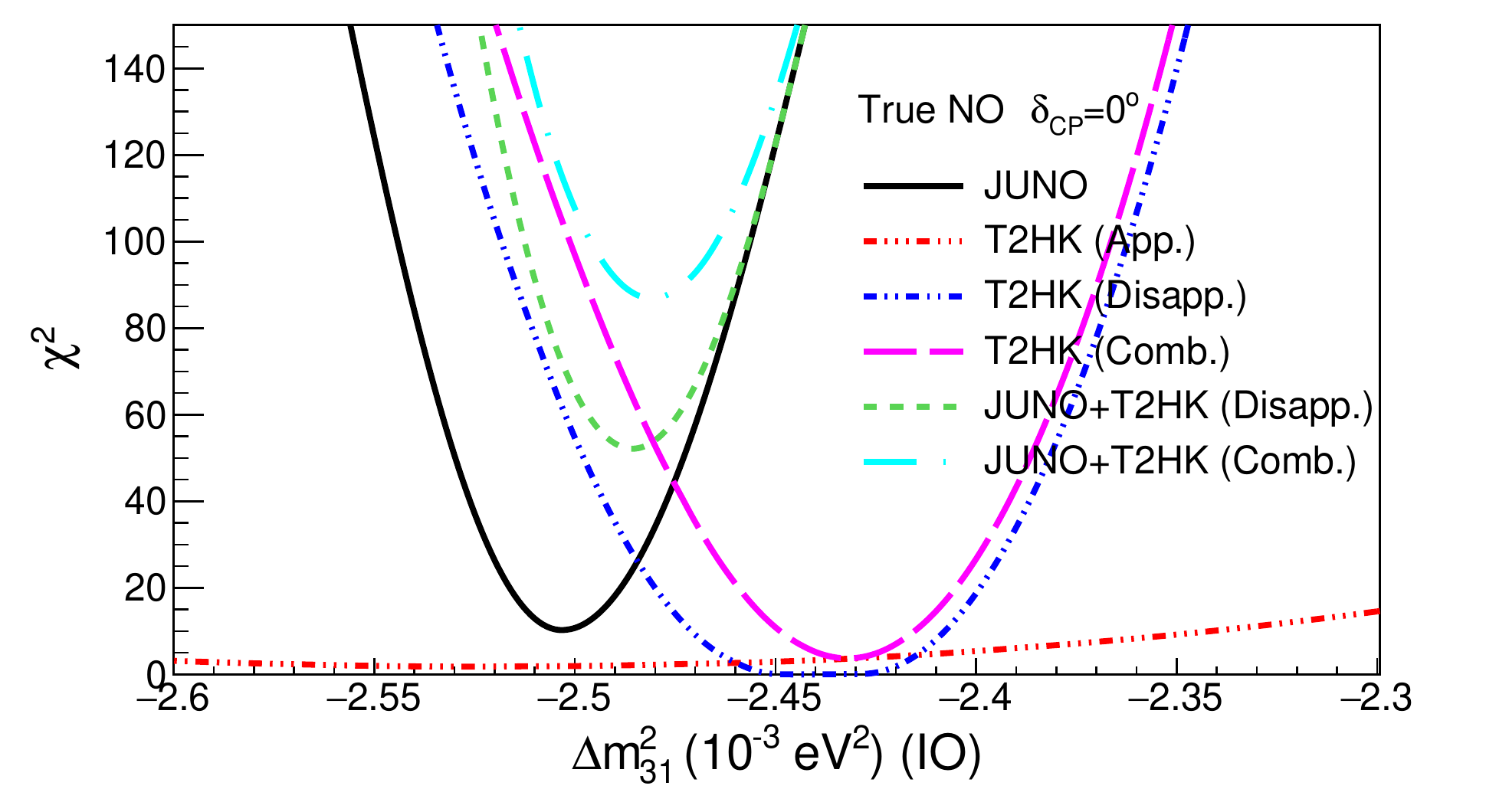}
 }
\qquad
\begin{minipage}{8cm}
\includegraphics[width=9cm]{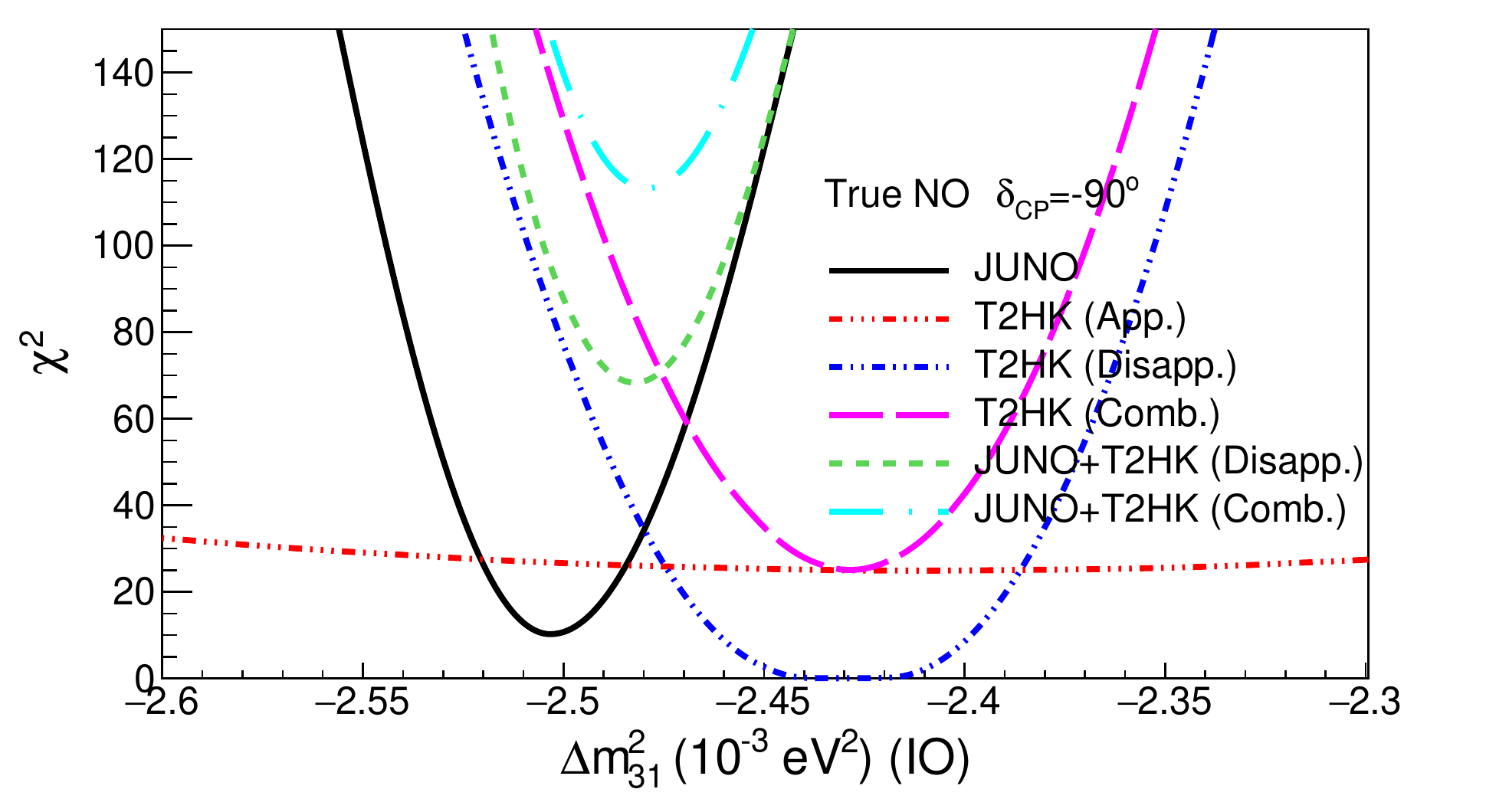}
\end{minipage}
\caption{Mass ordering sensitivity expected from a combined analysis of JUNO and T2HK, shown as a function of $\Delta m^2_{31}$(IO). Left panel is for $\delta_{\rm CP}=0^{\circ}$ and the right panel is for $\delta_{\rm CP}=-90^{\circ}$.}
\label{fig:junot2hk}
\end{figure*}

Finally, we do joint analysis of JUNO and T2HK and present our results in Fig.~\ref{fig:junot2hk}. Results are shown in two panels for two cases of $\delta_{\rm CP}=0^\circ$ and $-90^\circ$ motivated by the recent best-fit values of this parameter by NO$\nu$A \cite{con_talk} and T2K \cite{con_talk_t2k} respectively. Note that JUNO is independent of both $\theta_{23}$ as well as $\delta_{\rm CP}$. 

\begin{table}[h!]
\centering
\begin{tabular}{|c| c | c | c |c |}
\hline
& JUNO & T2HK   & JUNO+T2HK & True value \\
\hline 
$\delta_{\rm CP}=0^\circ$ & -2.503 &  -2.432 &  -2.481 & 2.531\\
\hline
$\delta_{\rm CP}=-90^\circ$ & -2.503 &  -2.429 &  -2.479 & 2.531\\
\hline
\end{tabular}
\caption{Values of $\Delta m_{31}^{2}$(IO) (in units of $10^{-3}$ eV$^{2}$), for which we get mass ordering $\chi^2$ minimum. 
}
\label{tab:junot2hkbf}
\end{table}
 \begin{table}[h!]
\centering
\begin{tabular}{|c| c | c | c |c | c |c |}
\hline
NO & $\chi^{2}_{J}$ & $\chi^{2}_{T}$ & $\chi^{2}_{J}+\chi^{2}_{T}$ & $\chi^{2}_{J+T}$ & $\%$ increase \\
\hline 
$\delta_{\rm CP}=0^\circ$ & 10.23 &  3.70 &  13.93 &  86.87 &  523 \\
\hline
$\delta_{\rm CP}=-90^\circ$ & 10.23 &  25.07 &  35.30 &  113.22 &  220\\
\hline
\end{tabular}
\caption{Mass ordering $\chi^2$ for true NO and test IO. We label the $\chi^2$ for JUNO and T2HK as $\chi^2_J$ and $\chi^2_T$, respectively. 
}
\label{tab:junot2hkchi}
\end{table}

We summarize in Tables~\ref{tab:junot2hkbf} and \ref{tab:junot2hkchi} the main results of this analysis. The values of $\Delta m_{31}^2$(IO) for which we get the minimum $\chi^2$ for JUNO and T2HK individually, as well as for JUNO and T2HK combined analysis are given in Table~\ref{tab:junot2hkbf}. In Fig.~\ref{fig:junot2hk}, we show the MO sensitivity as a function of $\Delta m^2_{31}$ (IO). This figure shows the sensitivity of each oscillation channel in JUNO and T2HK to MO, as well as the synergy between them. We can see from the Fig.~\ref{fig:junot2hk} and Table~\ref{tab:junot2hkbf} that even though data in both experiments were simulated at the same assumed true value of $\Delta m_{31}^2$(NO), the best-fit values of $\Delta m_{31}^2$(IO) comes out to be different for JUNO and T2HK. As a result when we perform a combined analysis of the two data sets, the expected MO sensitivity is significantly enhanced. While a simple sum of the $\chi^2$ for JUNO and T2HK gives 13.93 (35.3) for $\delta_{\rm CP}=0^\circ$ ($-90^\circ$), the combined analysis gives $\chi^2$ of 86.87 (113.22). This is a staggering synergistic increase in the sensitivity, shown in the last column of Table \ref{tab:junot2hkchi}. In particular, for true $\delta_{\rm CP}=0^\circ$, T2HK by itself is expected to not even return a $\chi^2_{T}=4$ for mass ordering \cite{Abe:2016ero}. A simple sum of the $\chi^2$s of T2HK and JUNO ($\chi^2_{J}$) would still not achieve $\chi^2=16$ sensitivity. However, a combined analysis of JUNO and T2HK gives $\chi^2_{J+T}\simeq 87$, which shows a whopping improvement of 523\%. For the $\delta_{\rm CP}=-90^\circ$ case, both T2HK and JUNO individually give a good sensitivity to mass ordering, however, even in this case doing a combined analysis would lead to mass ordering being discovered with a much higher sensitivity. Let us mention that for T2HK, the mass ordering sensitivity is very poor for $\delta_{\rm CP} = 0^\circ$ as compared to $\delta_{\rm CP} = -90^\circ$ in normal ordering because of the hierarchy - $\delta_{\rm CP}$ degeneracy \cite{Barger:2001yr,Ghosh:2015ena,Prakash:2012az}. Indeed for a large fraction of $\delta_{\rm CP}$ values, the MO $\chi^2$ for T2HK is below 4. We see from our results above that combining the JUNO data to T2HK's can alleviate this problem and we expect greater than $9\sigma$ MO sensitivity for all values of $\delta_{\rm CP}$.

For the sake of understanding the underlying physics, we also show the MO sensitivity of the appearance and the disappearance channels in T2HK separately in Fig.~\ref{fig:junot2hk}. We can see that while the appearance channel alone has MO sensitivity, its $\Delta m_{31}^2$ dependence is very shallow. On the other hand, the disappearance channel alone does not have any MO sensitivity, however, it has a very sharp dependence on $\Delta m_{31}^2$. As a result, the combined $\chi^2$ to MO from JUNO and the disappearance channel of alone of T2HK has a very strong synergy coming from their strong $\Delta m_{31}^2$ dependence. This can been seen in the green dashed curve of Fig.~\ref{fig:junot2hk}. Addition of the appearance channel improves the sensitivity further, giving the highest possible sensitivity coming from the combination of these two experiments. 

It is pertinent to compare the MO sensitivity of the combined JUNO and T2HK set-up with the sensitivity of some of the most promising forthcoming experiments such as DUNE \cite{DUNE:2020jqi}, KM3NeT-ORCA \cite{KM3Net:2016zxf} and PINGU \cite{IceCube:2016xxt}. The future accelerator based experiment DUNE can measure MO with at least $10 \sigma$ ($16 \sigma$) C.L., for $\delta_{\rm CP} = 0^\circ$ ($-90^\circ$) in its 7 years of running irrespective of the true values of $\theta_{23}$. On the other hand, the analysis with the atmospheric neutrinos at the KM3NeT facility can provide a measurement of MO at $4 \sigma$ after 5 years of running for $\delta_{\rm CP} = 0^\circ$ and $\theta_{23} = 42^\circ$. For PINGU, which is a proposed low-energy extension to the IceCube can measure MO by studying the atmospheric neutrinos with a significance of at least $3 \sigma$ for $\theta_{23} = 45^\circ$ when a 68\% uncertainty on the other oscillation parameters are considered in its 4 year running. From the above discussion we understand that the combination of T2HK and JUNO outperforms KM3NeT-ORCA and PINGU. Regarding DUNE, the sensitivity of T2HK+JUNO is comparable to DUNE for $\delta_{\rm CP} = 0^\circ$ but DUNE outperforms T2HK+JUNO for $\delta_{\rm CP} = -90^\circ$.

\section{Remedying the energy resolution challenge for JUNO}

\begin{figure}
\hspace{-0.38 in}
 %   \centering
    \includegraphics[width=0.53\textwidth]{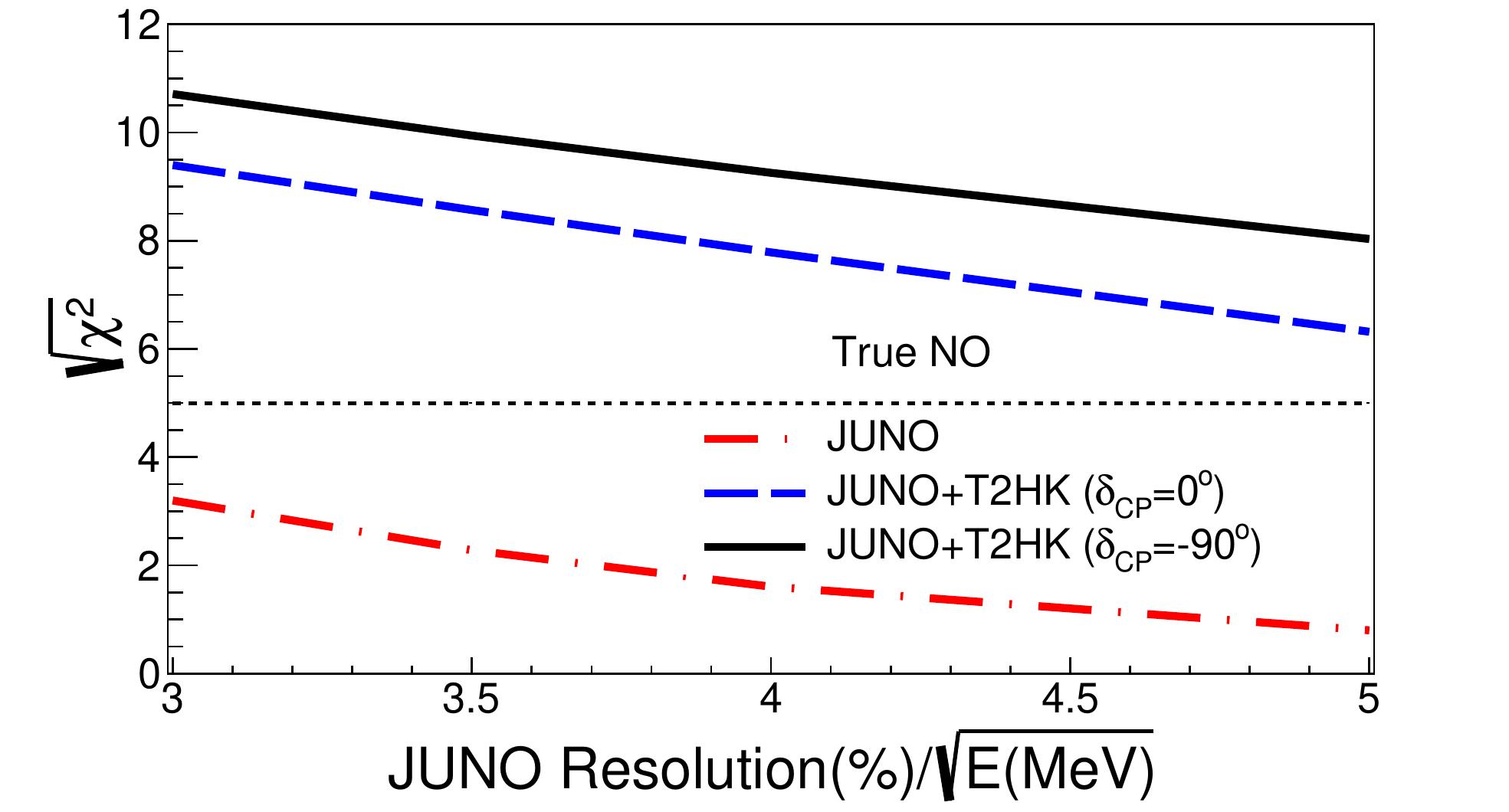}
    \caption{Expected MO sensitivity $\sqrt{\chi^{2}}$ as a function of energy resolution of JUNO.}
    \label{juno_res}
\end{figure}

In order to achieve $3\sigma$ sensitivity for MO, the JUNO detector will need better than 3\% energy resolution. This is unprecedented, challenging, as well as expensive. JUNO sensitivity to MO falls sharply with worsening of the energy resolution and expected to go to below 1 $\sigma$ for 5$\%$ energy resolution. Adding T2HK and JUNO and doing a combined analysis can circumvent this challenge. We show in Fig.~\ref{juno_res} the $\sqrt{\chi^2}$ of the combined analysis of T2HK and JUNO as a function of the energy resolution in JUNO. %{\color{red} Here we have taken the combined appearance and disappearance $\chi^2$ of T2HK}  
We show the results for both $\delta_{\rm CP}=0^\circ$ and $-90^\circ$. We can see that even with 5\% energy resolution in JUNO, we hope to get MO sensitivity that is well above $6\sigma$ for $\delta_{\rm CP}=0^\circ$. This would further increase  to 8$\sigma$ if $\delta_{\rm CP}=-90^{\circ}$.

\section{Conclusion}%

Measuring the MO is one of the most important aspects of neutrino physics. Ideally we want at least $5\sigma$ sensitivity to confirm the correct MO. JUNO experiment is being built to determine the MO, however, it is expected that even to achieve about 3-4 $\sigma$ sensitivity one would need better than 3\% energy resolution in JUNO. On the other hand, T2HK being built for CP studies, has a rather poor MO sensitivity for a large range of $\delta_{\rm CP}$ values. This compromises its CP sensitivity. In this article, we have shown that there is synergy between these two experiments due to the difference in which their oscillation probabilities depend on $|\Delta m_{31}^2|$. This synergy results in a staggering increase of the expected MO sensitivity when we perform a joint analysis. We showed that this increase could be between 220\% to 520\% depending on the true value of $\delta_{\rm CP}$. We also showed how this synergy can be instrumental in alleviating the energy resolution challenge for JUNO. We showed that even with 5\% energy resolution in JUNO, the combined analysis gives an expected MO sensitivity of greater than $6\sigma$. 

To summarize, we have shown that due to synergy coming from the $|\Delta m_{31}^2|$ dependence, MO sensitivity of greater than $9\sigma$ can be achieved by the combined analysis of JUNO and T2HK. It would be interesting to see if JUNO can achieve 3$\%$ energy resolution. We have shown that even if JUNO fails to achieve that, we could still get better than $6\sigma$ MO sensitivity from a joint analysis of JUNO with T2HK. Hence, the energy resolution challenge for JUNO is not even needed and the experiment can go ahead with a detector of 5\% energy resolution.

\begin{acknowledgments}
 The authors would like to thank Suprabh Prakash for providing the .glb file for JUNO and useful discussions regarding the JUNO experiment. MG acknowledges Ramanujan Fellowship of SERB, Govt. of India, through grant no: RJF/2020/000082. This project has received funding/support from the European Union’s Horizon 2020 research and innovation programme under the Marie Skłodowska -Curie grant agreement No 860881-HIDDeN.
\end{acknowledgments}

\bibliography{t2hk_juno_hier}% Produces the bibliography via BibTeX.

\end{document}